\documentstyle[12pt]{article}
\begin{document}
\textwidth 19 true cm
\textheight 23 true cm
\baselineskip=12pt
\voffset=-32 mm
\renewcommand{\baselinestretch}{1.5}
\renewcommand{\theequation}{\arabic{equation}}

\title{{Surface Polymer Network Model and}
\\{Effective Membrane Curvature Elasticity}}
\author{ R.Podgornik\thanks{On leave from J.Stefan Institute, P.O.B. 100,
61000 Ljubljana, SLOVENIA.}
\\{Laboratory of Structural Biology}
\\{Division of Computer Research and Technology}
\\{National Institutes of Health, Bethesda, MD 20892} }

\date{}
\begin{titlepage}
\maketitle
\begin{abstract}

\baselineskip=10pt

A microscopic model of a surface polymer network - membrane system is
introduced, with contact polymer surface interactions that can be
either repulsive or attractive and sliplinks of functionality four
randomly distributed over the supporting membrane surface anchoring
the polymers to it. For the supporting surface perturbed from a planar
configuration and a small relative number of surface sliplinks, we
investigate an expansion of the free energy in terms of the local
curvatures of the surface and the surface density of sliplinks,
obtained through the application of the Balian - Bloch - Duplantier
multiple surface scattering method. As a result, the dependence of the
curvature elastic modulus, the Gaussian modulus as well as
of the spontaneous curvature of the "dressed" membrane, ~{\sl i.e.}
polymer network plus membrane matrix, is obtained on the mean polymer
bulk end to end separation and the surface density of sliplinks.

\vskip 0.5 cm

{\obeylines
}

\vskip 0.5 cm

\centerline{\today }

\end{abstract}
\end{titlepage}

\textwidth 19 true cm
\textheight 23 true cm
\baselineskip=15pt
\parskip=15pt plus 1pt 
\parindent=40pt

\section{Introduction}

Polymer networks often provide highly specialized elastic properties
to biological systems.  Structure of the red blood cell (RBC) membrane
is typical in this respect \cite{evans} in the sense that it unites the lipid
matrix and the (polymer) spectrin network into a single "dressed"
membrane, where the polymer network not only interacts with the
underlying lipid surface but is actually anchored to it {\sl via}
ankyrin molecules.  The polymer network consists of negatively charged
spectrin tetramers of $\sim 200$ nm contour length, with intrinsic
persistence length of $\sim 10$ nm. The replicating network of
spectrin tetramers has a junction functionality of between $4$ and
$6$. The network junctions do not coincide with attachment
points (ankyrin molecules). The contour length of the spectrin
tetramers is between $2-6$ times larger then the separation between
anchoring points and the whole system thus looks like a crosslinked
two dimensional gel where the crosslinks are bound to remain on the
supporting lipid surface while the polymers are allowed to stretch
into the cytoplasmic (polymer rich) solution and away from the lipid
surface \cite{ref4}.

The problem of elastic properties of a ``dressed'' membrane, where the
individual polymer molecules or their junctions are confined to lie on
(or are embedded in) the supporting surface, motivated by our present
understanding of the RBC spectrin network, received recently a lot of
attention \cite{boal,everaers,kozlov}. The method of choice in most of
these investigations has been the Monte Carlo simulations that led to
several important insights regarding surface - tethered polymers
and/or surface polymer networks. The analytical limits for these
systems have been much harder to come by due to the complicated nature
of the polymer network - surface interactions. In this paper we shall
try to fill this gap and establish some approximate limiting
analytical results for the elastic properties of a surface polymer
network pinned to a supporting flexible membrane surface. More
specifically we shall investigate the modifications in the elastic
energy parameters of a flexible membrane wrought by the presence of a
surface polymer gel where the crosslinks (in this case presumed to
be sliplinks) are constrained to lie on the supporting membrane
surface. The parts of the polymeric chains between the junctions are
allowed to sample all the configurations on one side of the supporting
membrane, {\sl i.e.} are constrained to lie in the halfspace defined
by the position of the membrane surface (see Fig.1).

Several theoretical studies lately addressed the issues pertaining
to the modifications of elastic properties of a membrane in
contact with a polymer solution that interacts with it
\cite{degennes,ref2,ref1}. Though the problem in this case is simpler than in
the surface polymer gel case, it appears that certain features of
the approach employed \cite{ref1} can be generalized to this case.
Specifically we shall derive the curvature expansion of the free
energy of a surface polymer gel with deformed supporting surface and
show that the spontaneous curvature as well as the elastic modulus of
the "dressed" membrane become functions of the polymer parameters,
most notably the monomer - surface interaction and the polymer bulk
end - end separation.

We presume that the chains in between the junctions are Gaussian and
that their interaction with the supporting membrane surface, aside
from the sliplinks where they are pinned to the surface, is of a
contact type, thus simulating the screened electrostatic or other
short range interactions between polymer segments and the membrane
surface (see Fig.1). The proposed formulation of the surface gel on a
non - planar supporting surface problem is closely related to the
formulation of similar polymer problems in the bulk introduced by
Edwards and Freed
\cite{ref5} and is particularly suited for the application of the
Balian - Bloch - Duplantier \cite{ref6} multiple surface scattering
Green function formalism, which gives a local curvature expansion for
a Helmholtz equation in a region with deformed boundaries. 

A major drawback of this approach is that we have been unable to
include volume interactions between polymer chains, extensively
studied in \cite{boal,everaers}, into the formulation in a consistent
manner. However, this drawback is substantially counterbalanced by the
fact that first of all analytic results are feasible in this
approximation and that the effect of even very complicated polymer -
surface interactions, as is the case in this model system, can be
readily investigated within the developed formalism.

\section{Model}

We define our system as a statistical ensemble of freely mobile sliplinks
confined to the supporting surface ${\bf r}_{\alpha}$ (here
and below the Greek indices stand for coordinates along the supporting
surface) produced after $N$ chains ${\cal N}$ monomers long are crosslinked $M$ 
times with functionality four. Following \cite{ref5} the crosslinks
are taken to be introduced by an external auxiliary field, $\phi({\bf
r})$, that produces polymer chains with Green function ${\cal
G}_{\phi}({\bf R},{\bf R}';{\cal N})$ if the chains are $\cal N$
segments long. One can show through a diagrammatic expansion of this
Green function that if it is averaged over a Gaussian distribution of $\phi({\bf
r})$ this operation formally introduces an arbitrary number of
cross-links of functionality four, distributed randomly through space
and along the chains, with the partition function 
\begin{equation}
\Xi(N,{\cal N}) = \left<\left(\int\!\!\int d^3{\bf R}{\cal G}_{\phi}({\bf R},{\bf
R}';{\cal N})d^3{\bf R}'\right)^N\right>_{\phi},
\end{equation}
for $N$ chains.

If one now expands the Green function in terms of $\sqrt{\mu} \phi({\bf r})$
and then averages with $ \left<\dots \right>_{\phi} $, the coefficient
of $\mu^M$ contains all the terms with $M$ crosslinks randomly
distributed across the space and along the chains. Thus the partition
function of a system with $M$ crosslinks can be formally written as
\begin{equation}
\Xi(M,N,{\cal N}) = \frac{1}{2\pi \imath }\oint_{\cal C}
\frac{M!~d\mu}{\mu^{M+1}} \left<\left(\int\!\!\int d^3{\bf 
R}{\cal G}_{\phi}({\bf R},{\bf R}';{\cal
N})d^3{\bf R}'\right)^N\right>_{\phi}.
\end{equation}

For the system discussed here the partition function can be obtained
along the lines of \cite{ref5}, except that the auxiliary field should
now be confined to the surface ${\bf r}_{\alpha}$. This would signify
that it should effect not the equation defining the Green function
itself, but rather its boundary condition at this surface. Also if
there exist other interactions between the polymers and the surface,
they should also show up only in the boundary condition if their range is
short enough.

Putting now everything together we derive the partition function of
our model system in the following compact notation
\begin{equation}
\Xi (M,N, {\cal N}) = {{M!}\over{2\pi \imath}} \oint_{\cal C} {{d\mu}\over{\mu^{M+1}}}
\left< \left[ {\cal L}^{-1}\!\!\! \int\!\!\!\int\!\! d^3{\bf r}d^3{\bf r}' {\cal G}_{\phi}( 
{\bf r},{\bf r}';s) \right]^N \right>_{\phi} = {{M!}\over{2\pi \imath}} \oint_{\cal C} {{d\mu}\over{\mu^{M+1}}}
\left< \Xi_{\phi} (\mu ,{\cal N})^N \right>_{\phi}, 
\label{equ1}
\end{equation}
where ${\cal G}_{\phi}({\bf r},{\bf r}';s) $ is the Laplace transformed 
Green function of a free flight polymer, satisfying a Helmholtz - type
equation \cite{ref0}
\begin{equation}
\left[ \nabla^2 - s \right]{\cal G}_{\phi}({\bf r},{\bf r}';s) = -\delta^3 (
{\bf r} - {\bf r}' ).
\label{equ2}
\end{equation}
Here we made a transformation ${\bf r} \to {{\sqrt{6}{\bf
r}}\over{\ell}}$, measuring the spatial dimensions in units of
${{\ell}\over{\sqrt{6}}}$, where $\ell$ is the steplength. ${\cal
L}^{-1}$ stands for the inverse Laplace transform, {\sl i.e.} ${\cal
L}^{-1}f(s) = \oint_{\cal C} e^{s{\cal N}} f(s) ds = f({\cal N})$ and
the definition of $\Xi_{\phi} (\mu ,{\cal N})$ is obvious. At the
supporting membrane surface ${\bf r}_{\alpha}$ the Green function
satisfies \cite{eisenriegler}
\begin{equation}
{{\partial {\cal G}_{\phi}({\bf r}_{\alpha},{\bf r}';s)}\over
{\partial {\bf n}_{\alpha}}} - \kappa {\cal G}_{\phi}({\bf
r}_{\alpha},{\bf r}';s) + \sqrt{\mu} 
\phi ({\bf r}_{\alpha}){\cal G}_{\phi}({\bf r}_{\alpha},{\bf r}';s) = 0,
\label{equ3}
\end{equation}
where $\kappa$ is the strength of the contact membrane surface -
polymer interaction. It can be either repulsive ($\kappa > 0$) or
attractive ($\kappa < 0$). The last term in the above equation,
proportional to $\sqrt{\mu}$, creates at random positions along the
supporting surface ${\bf r}_{\alpha}$ as well as along the polymer
chains junctions of fourfold functionality (sliplinks), of chemical
potential $\mu$ per single junction, after the Green function is
averaged over a Gaussian distribution of the auxiliary field $\phi
({\bf r}_{\alpha})$
\cite{ref5},
\begin{equation}
\left< \dots \right>_{\phi} = \int {\cal D}\phi ({\bf r}_{\alpha}) 
\left( \dots \right) e^{-{\textstyle\frac{1}{2}} \int \phi^2 ({\bf r}_{\alpha}) 
d^2{\bf r}_{\alpha}} / \int {\cal D}\phi ({\bf r}_{\alpha}) 
e^{-{\textstyle\frac{1}{2}} \int \phi^2 ({\bf r}_{\alpha}) d^2{\bf r}_{\alpha}}.
\label{equ4}
\end{equation}
The averaging over a Gaussian distribution of the auxiliary field
preserves only terms with even powers of $\sqrt{\mu}$.
 
We shall now try to obtain an approximate solution of the above model
in the limit of ${\textstyle\frac{M}{N}} \ll 1$ while separately $N \gg 1$ and $M
\gg 1$, with the additional proviso that the volume (or surface)
density of chains $\frac{N}{V} \left( {\rm or}
~\frac{N}{S_{\omega}}\right.$ depending on the sign of $\left. \kappa \right)$
and the surface density of sliplinks ${\textstyle\frac{M}{S_{\omega}}}$ are finite
(thermodynamic limit).

We first of all cast the model in a form that will be appropriate for
later formal developements. It is easiest to obtain a power series
solution of Eq.\ref{equ3} in terms of the free space solution ${\cal
G}_0({\bf r},{\bf r}';s)$ {\sl i.e.} 
\begin{equation}
{\cal G}_0({\bf r},{\bf r}';s) = \frac{e^{-\sqrt{s}\vert {\bf r} -
{\bf r}'\vert}}{4\pi \vert {\bf r} - {\bf r}'\vert}.
\label{green}
\end{equation}
This series has been already derived by Balian and Bloch in
\cite{ref6} and we merely quote their result. Introducing $\Gamma ({\bf
r}_{\gamma},{\bf r}_{\epsilon};s)$ as
\begin{equation}
\Gamma ({\bf r}_{\alpha},{\bf r}_{\beta};s) + 2\kappa\int dS_{\gamma}
{\cal G}_0({\bf r}_{\alpha},{\bf r}_{\gamma};s)
\Gamma ({\bf r}_{\gamma},{\bf r}_{\beta};s) = \delta^2 ({\bf r}_{\alpha} -
{\bf r}_{\beta}),
\label{equ10}
\end{equation}
one obtains the following expansion for the Green function
\begin{eqnarray}
\kern-80pt {\cal G}_{\phi}({\bf r},{\bf r}';s) &=& {\cal G}_0({\bf r},{\bf r};s) + 2
\int\!\!\int dS_{\alpha} dS_{\beta} {\cal G}_0({\bf r},{\bf
r}_{\alpha};s) \Gamma ({\bf r}_{\alpha},{\bf r}_{\beta};s)
\nabla_{\beta} {\cal G}_0({\bf r}_{\beta},{\bf r}';s) + \nonumber\\
&+& 2^2 \int\!\!\!\!\int\!\!\!\!\int\!\!\!\!\int dS_{\alpha}
dS_{\beta} dS_{\gamma} 
dS_{\delta} {\cal G}_0({\bf r},{\bf 
r}_{\alpha};s) \Gamma ({\bf r}_{\alpha},{\bf r}_{\delta};s)
\frac{\partial {\cal G}_0({\bf r}_{\delta},{\bf r}_{\gamma};s)
}{\partial {\bf n}_{\delta}} \Gamma ({\bf r}_{\delta},{\bf r}_{\beta};s)
\nabla_{\beta} {\cal G}_0({\bf r}_{\beta},{\bf r}';s) + \dots 
\nonumber\\
~
\label{balian}
\end{eqnarray}
where
\begin{equation}
\nabla_{\alpha} = {{\partial}\over{\partial {\bf n}_{\alpha}}} - \kappa +
\sqrt{\mu} \phi ({\bf r}_{\alpha}).
\end{equation}
We might just add here that for a planar surface the normal derivative
of the Green function, $\frac{\partial {\cal G}_0({\bf r}_{\delta},{\bf r}_{\gamma};s)
}{\partial {\bf n}_{\delta}}$, is zero and thus the expansion
Eq.\ref{balian} is basically an expansion in the curvature of the
bounding surface.

Combining now both Eqs.\ref{equ10} and \ref{balian} we can derive
the partition function in a rather transparent form
\begin{eqnarray}
\Xi_{\phi} (\mu ,{\cal N}) &=& \Xi_b({\cal N}) + 2~Tr~U_{\phi}({\bf
r}_{\alpha},{\bf r}_{\beta};  
{\cal N}) = \nonumber\\
&=& V + 2\int dS_{\omega} U_{\phi}({\bf r}_{\omega},{\bf r}_{\omega};
{\cal N}),
\label{equ5}
\end{eqnarray}
where $\Xi_b({\cal N}) = \int\!\!\!\int d^3{\bf r}d^3{\bf r}' {\cal
G}_0( {\bf r},{\bf r}';{\cal N}) = V$ is the volume of the polymer
solution with unperturbed boundaries and the operator $U_{\phi}({\bf
r}_{\alpha},{\bf r}_{\beta};s)$ is obtained as a solution of
\begin{equation}
U_{\phi}({\bf r}_{\alpha},{\bf r}_{\beta};s) - 2 \int dS_{\gamma}
U_{\phi}({\bf r}_{\alpha},{\bf r}_{\gamma};s) \nabla_{\gamma}{\cal G}_0(
{\bf r}_{\gamma},{\bf r}_{\beta};s) = \nabla_{\alpha} F({\bf r}_{\alpha},
{\bf r}_{\beta};s),
\label{equ6}
\end{equation}
with $dS_{\omega}$ being the area element of the surface ${\bf 
r}_{\alpha}$, and  
\begin{equation}
F({\bf r}_{\alpha},{\bf r}_{\beta};s) = \int\!\!\!\int d^3{\bf r}d^3{\bf r}'
{\cal G}_0({\bf r},{\bf r}_{\alpha};s){\cal G}_0({\bf r}_{\beta},{\bf r}';s).
\label{equ7}
\end{equation}
Eqs.\ref{equ5} and \ref{equ6} represent just a more compact
formalisation of the results already derived in \cite{ref6} that is
conveniently suited for our purposes.

Formally the above equations define our model system. Their solution
is obviously non-trivial and additional approximations have to be
introduced in order to make the problem tractable. First of all we
shall introduce the approxiation of low polymer density that will make
the evaluation of ${\phi}({\bf r})$ average tractable in the form of a
power series in the chemical potential of the sliplinks, $\mu$. Next
we shall resolve the integration over $\mu$ by reverting to the saddle 
- point {\sl ansatz} and finally we shall use the approximation of the
local tangential plane that explicitely introduces the surface
curvature dependence of the different Green functions to evaluate the
different curvature terms in the final expression for the partition
function. 

\section{Analysis}

Starting from Eq.\ref{equ5}, $\Xi_{\phi} (\mu ,{\cal N})^N$ can be written
in the form of a "virial" expansion in terms of the polymer volume density. 
If this density is small enough we can stop at the first order 
term thus obtaining
\begin{equation}
\Xi (M,N, {\cal N}) \cong {{M!}\over{2\pi \imath}} \oint_{\cal C} 
{{d\mu}\over{\mu^{M+1}}}~e^{N\ln{\left< \Xi_{\phi} (\mu ,{\cal N})
\right>_{\phi} }} = {{M!}\over{2\pi \imath}} \oint_{\cal C} 
{{d\mu}\over{\mu^{M+1}}}~e^{N\ln{\left( V + 2~Tr~\left< U_{\phi}(
{\bf r}_{\alpha},{\bf r}_{\beta};{\cal N}) \right>_{\phi} \right) }}.
\label{equ7b}
\end{equation}

As we have done before \cite{ref1} we now substantially reduce the
heavy algebra proceeding from the above equations if we presume that $
\vert \kappa \vert \gg {{\partial \log{{\cal G}_0({\bf r}_{\alpha},
{\bf r}';s) }}\over{\partial {\bf n}_{\alpha}}}$. First of all we
solve Eq.\ref{equ6} in the form of a formal series in $\phi ({\bf
r}_{\alpha})$.  This series solution of Eq.\ref{equ6} for $
U_{\phi}({\bf r}_{\alpha}, {\bf r}_{\beta};s) $ can be effectively
resummed and averaged over the Gaussian distribution of the auxiliary
field $\phi ({\bf r}_{\alpha})$ Eq.\ref{equ4}. Only even powers of $\phi ({\bf
r}_{\alpha})$ survive this averaging and we can regroup the remaining
terms obtaining a power series in $\mu$. By defining
\begin{eqnarray}
U_0 ({\bf r}_{\alpha},{\bf r}_{\beta};s) &=& -\kappa \int dS_{\gamma} F({\bf
r}_{\alpha},{\bf r}_{\gamma};s)\Gamma ({\bf r}_{\gamma}, {\bf
r}_{\beta};s) \nonumber\\ 
\Delta ({\bf r}_{\alpha},{\bf r}_{\beta};s) &=& \int dS_{\gamma}
{\cal G}_0({\bf r}_{\alpha},{\bf r}_{\gamma};s) \Gamma ({\bf r}_{\gamma},
{\bf r}_{\beta};s) \nonumber \\
W_0({\bf r}_{\alpha},{\bf r}_{\beta};s) &=& {\textstyle\frac{1}{2}}\int\!\!\!\int dS_{\gamma}
dS_{\epsilon} F({\bf r}_{\alpha},{\bf r}_{\gamma};s)\Gamma ({\bf r}_{\gamma},
{\bf r}_{\epsilon};s){\cal G}_0^{-1} ({\bf r}_{\epsilon},{\bf r}_{\beta};s)
\label{equ9}
\end{eqnarray}
we thus obtain the $\phi$ average of $U_{\phi}({\bf r}_{\alpha},{\bf
r}_{\beta};s)$ in the form of a series in the chemical potential of
the pinning sliplinks $\mu$
\begin{eqnarray}
\left< U_{\phi}({\bf r}_{\alpha},{\bf r}_{\beta};s)\right>_{\phi} &=&
U_0 ({\bf r}_{\alpha},{\bf r}_{\beta};s) \nonumber\\
&+& 4\mu \int dS_{\gamma} 
\Delta ({\bf r}_{\alpha},{\bf r}_{\gamma};s)W_0({\bf r}_{\gamma},
{\bf r}_{\beta};s) + \nonumber \\
&+& 16\mu^2 \int\!\!\!\int dS_{\gamma} dS_{\epsilon}
\Delta ({\bf r}_{\alpha},{\bf r}_{\epsilon};s)\Delta ({\bf r}_{\epsilon},
{\bf r}_{\gamma};s)W_0({\bf r}_{\gamma},{\bf r}_{\beta};s) + \dots ,\nonumber\\
~
\label{equ8}
\end{eqnarray} 
where we redefined the chemical potential $\mu \rightarrow \mu \Delta
( {\bf r}_{\alpha},{\bf r}_{\alpha};{\cal N})$, with $
\Delta ( {\bf r}_{\alpha} \-, {\bf r}_{\alpha} ;{\cal N}) \to 
\pi \sqrt{\cal N}V^{-{\textstyle\frac{1}{3}}} \gg 1$. This relation stemms from the 
conservation of the degrees of freedom of the chain \cite{ref5}.  At
each order of the $\mu$ expansion we retain only the largest term
while omitting all the additional powers of $\Delta^{-1}( {\bf
r}_{\alpha},{\bf r}_{\alpha};{\cal N})$. The above expansion in terms
of $\mu$ is convergent if ${\textstyle\frac{M}{N}} \ll 1$, and we can stop at the
second order term.

The zero order term in this expansion, Eq.\ref{equ8}, is exactly the one
corresponding to the example treated previously \cite{ref1}, which
corresponds to the case without any surface pinning of the polymers
through the sliplinks.

The approximate form of $\left< \Xi_{\phi} (\mu ,{\cal N})\right>_{\phi} $ is
thus obtained as
\begin{equation}
\left< \Xi_{\phi} (\mu ,{\cal N})\right>_{\phi} 
= V + 2~\int dS_{\omega}\left< U_{\phi}({\bf r}_{\omega},{\bf r}_{\omega};{\cal N}) \right>_{\phi}
\cong \Xi_0({\cal N}) + \mu \Xi_1({\cal N}) + 
{\textstyle\frac{1}{2}} \mu^2 \Xi_2({\cal N}) + \dots ,
\label{equ10a}
\end{equation}
where aside from $V$ all the other $\mu$ terms in the expansion can be 
obtained by comparison with Eq.\ref{equ8} as
\begin{eqnarray}
\Xi_{0}({\cal N}) &=& \Xi_b({\cal N}) + 2~{\cal L}^{-1}\left[ {\rm Tr}~
U_0({\bf r}_{\alpha},{\bf r}_{\beta};s) \right] \nonumber\\
\Xi_1({\cal N}) &=& 8~{\cal L}^{-1}\left[ {\rm Tr} \int dS_{\gamma} \Delta
({\bf r}_{\alpha},{\bf r}_{\gamma};s) W_0 ({\bf r}_{\gamma},{\bf
r}_{\beta};s) \right] \nonumber\\
\Xi_2({\cal N}) &=& 64~{\cal L}^{-1}\left[ {\rm Tr} \int\!\!\int
dS_{\gamma}dS_{\epsilon} \Delta ({\bf r}_{\alpha},{\bf r}_{\gamma};s)
\Delta ({\bf r}_{\gamma},{\bf r}_{\epsilon};s) W_0 ({\bf
r}_{\epsilon},{\bf r}_{\beta};s) \right], \nonumber\\
~
\label{equxi}
\end{eqnarray}
where $\Xi_b({\cal N}) = {\cal L}^{-1}\left[ \frac{1}{s} \int d^3{\bf
r} \right] = V$.

The final $\mu$ integration in Eq.\ref{equ7b} is now of the form
\begin{equation}
\Xi (M,N,{\cal N}) \cong {{M!} \over{2\pi \imath}} \oint_{\cal C} {d\mu} e^{
N\ln{\left( \Xi_0({\cal N}) + \mu \Xi_1({\cal N}) + 
{\textstyle\frac{1}{2}} \mu^2 \Xi_2({\cal N}) + \dots\right) - (M+1)\ln{\mu} }},
\label{equ11}
\end{equation}
which is amenable to an analytic treatment through the saddle - point 
approximation in the thermodynamic limit of $N \gg 1$ and $M \gg 1$, 
yielding an approximate result $\mu = \mu^{\ast} ({\textstyle\frac{M}{N}})$, where to 
the lowest order we obtain $\mu^{\ast} \sim {\textstyle\frac{M}{N}} 
{\textstyle\frac{\Xi_0({\cal N})}{\Xi_1({\cal N})}} + \dots$. The saddle - point free 
energy ${\cal F}$ is thus obtained as ($\beta = (kT)^{-1}$) 
\begin{eqnarray}
\beta{\cal F} &=& -\ln{\Xi (M,N,{\cal N})} \cong \nonumber\\
&\cong& -N\ln{\left( \Xi_0({\cal N}) + \mu^{\ast} \Xi_1({\cal N}) + 
{\textstyle\frac{1}{2}} \mu^{\ast 2} \Xi_2({\cal N})\right) + (M+1)\ln{\mu^{\ast}} }
\label{equ11a}
\end{eqnarray}

Following the expansion of the statistical sum in terms of the
relative number of sliplinks, {\sl i.e.} ${\textstyle\frac{M}{N}}$, we now
additionally presume that the supporting surface ${\bf r}_{\alpha}$ is
deformed and we investigate the lowest order expansion of the free
energy in terms of the local curvatures of the deformed surface. We
approach this problem by the Balian - Bloch - Duplantier method
\cite{ref6} as was explained in detail for a somewhat simpler case of
polymer - surface interactions \cite{ref1}. 

The principle of the curvature expansion is simple. First of all we
note that the partition function Eq.\ref{equxi} contains different
surface integrals of either $G_0({\bf r}_{\alpha},{\bf r}_{\beta};s)$,
$\Delta ({\bf r}_{\alpha},{\bf r}_{\beta};s)$ or $F ({\bf
r}_{\alpha},{\bf r}_{\beta};s)$. Since all these quantities depend
only on the difference of the surface coordinates in the arguments we
can expand them around the local tangential plane up to the second
order in the deviations from that plane and then evaluate the surface
integrals explicitely.  This expansion in terms of the deviations from
the local tangential plane is valid only as long as the Green function
is of a range smaller than the local curvatures. Practically this
would mean that one can always find a range of curvature values where
the expansion of a Yukawa - type Green function Eq.\ref{green} in the
vicinity of a local tangential plane of a non-planar surface is
justified. 

We write the equation of the bounding surface ${\bf r}_{\alpha}$ in
the reference frame of the tangential plane at ${\bf r}_{\omega}$ as
\begin{equation}
z_{\omega}(x,y) \cong {\textstyle\frac{1}{2}}
\left(\frac{x^2}{R_{\omega_1}} + \frac{y^2}{R_{\omega_2}}\right) +
\dots
\end{equation}
where $R_{\omega_1}$ and $R_{\omega_2}$ are the two principal radii of
curvature at the point ${\bf r}_{\omega}$ on the bounding surface. 

We can now straightforwardly derive the form of the curvature
dependence up to and including second order of the Green function by
expanding the coordinate dependence in the vicinity of tangential
plane for each point along the bounding surface. The only non-zero terms in
the expansion of $G_0({\bf r}_{\alpha},{\bf r}_{\beta};s)$ are the
zero and the second order, thus giving
\begin{equation}
G_0({\bf r}_{\alpha},{\bf r}_{\beta};s)~ \cong~ \stackrel{(p)}{G_0}({\bf
r_{\alpha}},{\bf r_{\beta}};s) + {\textstyle{1\over 2}}
z_{\alpha}^2(x,y) {{\partial^2 \stackrel{(p)}{G_0}({\bf
r_{\alpha}},{\bf r_{\beta}};s)}\over{\partial z_{\alpha}^2}} + \dots
\label{equtop}
\end{equation}
where the superscript $p$ stands for 'planar' (i.e. zero curvature)
approximation of the superscripted quantity. Furthermore ${{\partial
G_0({\bf r_{\alpha}},{\bf r_{\beta}};s)}\over{\partial {\bf
n_{\alpha}}}}$ obviously has only odd terms in the expansion starting
with
\begin{equation}
{{\partial G_0({\bf r_{\alpha}},{\bf r_{\beta}};s)}\over{\partial {\bf
n_{\alpha}}}} \cong - \left({{\partial z_{\alpha}(x,y)}\over{\partial
x}}{{\partial}\over{\partial x}} + {{\partial z_{\alpha}(x,y)}\over{\partial
y}}{{\partial}\over{\partial y}}\right)\stackrel{(p)}{G_0}({\bf
r_{\alpha}},{\bf r_{\beta}};s) + \dots
\label{equbot}
\end{equation}
since the local normal is defined as ${\bf n}_{\alpha} =
(1,-{{\partial z_{\alpha}(x,y)}\over{\partial x}},-{{\partial
z_{\alpha}(x,y)}\over{\partial y}})$. The third origin of the
curvature dependence is the surface area element $dS_{\omega}$ itself
in Eqs.\ref{equ10} and \ref{equ9}
\begin{equation}
dS_{\omega} \cong d\stackrel{(p)}{S}_{\omega} +
~{\textstyle\frac{1}{2}}\left( \nabla z_{\omega}(x,y) \right)^2
d\stackrel{(p)}{S}_{\omega}. 
\label{equzad}
\end{equation}
Finally expressions Eqs.\ref{equtop}, \ref{equbot} and \ref{equzad}
have to be averaged locally over all the directions of the principal
curvature axes.

These fundamental dependencies of the Green function
$G_0({\bf r}_{\alpha},{\bf r}_{\beta};s)$ and its normal derivative on
the local curvature now generate the corresponding functional
dependence for the surface Green function $\Gamma ({\bf
r_{\alpha}},{\bf r_{\beta}};s)$ through the defining equation
Eq.\ref{equ10} and for $F({\bf r}_{\alpha},{\bf r}_{\beta};s)$ that
can be written in an alternative form
\begin{eqnarray}
F({\bf r_{\alpha}},{\bf r_{\beta}};s) &=& \int\!\!\!\int d^3{\bf
r}d^3{\bf r'} G_0({\bf r},{\bf r_{\alpha}};s) G_0({\bf
r_{\beta}},{\bf r'};s) = \nonumber\\
&=& {1\over {s^2}}\left[{1\over 4} -
\int{{\partial G_0({\bf r_{\gamma}},{\bf
r_{\alpha}};s)}\over{\partial {\bf n_{\gamma}}}}dS_{\gamma} + \int\!\!\!\int
{{\partial G_0({\bf r_{\gamma}},{\bf
r_{\alpha}};s)}\over{\partial {\bf n_{\gamma}}}}{{\partial G_0({\bf
r_{\gamma '}},{\bf r_{\beta}};s)}\over{\partial {\bf n_{\gamma
'}}}} dS_{\gamma}dS_{\gamma '}\right] \nonumber\\
~
\label{equtra}
\end{eqnarray}
where the curvature dependence transpires more directly. With all
these provisos we can now evaluate the surface integrals in the
defining equations Eq.\ref{equ9} explicitely order by order in the inverse
curvature \cite{ref1}. Though this procedure is tedious it is
nevertheless straightforward. The following relations are obtained 
\begin{eqnarray}
\int \stackrel{(0)}{U}({\bf r_{\omega}},{\bf r_{\alpha}};s)
~d\stackrel{(p)}{S}_{\alpha}  & & = 
-{{\kappa}\over{4\sqrt{s}^3(\kappa + \sqrt{s})}} \nonumber\\
\int \stackrel{(1)}{U}({\bf r_{\omega}},{\bf r_{\alpha}};s)
~d\stackrel{(p)}{S}_{\alpha} & & = 
{{\kappa}\over{2 s^2(\kappa + \sqrt{s})}}\times \frac{1}{R_{\omega}} \nonumber\\
\int \stackrel{(2)}{U}({\bf r_{\omega}},{\bf r_{\alpha}};s)
~d\stackrel{(p)}{S}_{\alpha} & & =  
-{{\kappa}\over{4 \sqrt{s}^5(\kappa + \sqrt{s})}}\times \left[
{1\over{R_{\omega}^2}} - {{\kappa }\over{2
(\kappa + \sqrt{s})}}\left({1\over{R_{\omega}^2}} -
{1\over{R_{\omega_1}R_{\omega_2}}}\right)\right] \nonumber\\
\int \stackrel{(0)}{\Delta}({\bf r_{\omega}},{\bf r_{\alpha}};s)
~d\stackrel{(p)}{S}_{\alpha} & & = 
\frac{1}{2(\kappa + \sqrt{s})} \nonumber\\
\int \stackrel{(2)}{\Delta}({\bf r_{\omega}},{\bf r_{\alpha}};s)
~d\stackrel{(p)}{S}_{\alpha} & & =
\frac{1}{4\sqrt{s}(\kappa + \sqrt{s})}\times \left(
\frac{1}{R_{\omega}^2} - \frac{1}{R_{\omega_1}R_{\omega_2}}\right)
\nonumber\\ 
\int \stackrel{(0)}{W_0}({\bf r_{\omega}},{\bf r_{\alpha}};s)
~d\stackrel{(p)}{S}_{\alpha} & & =
\frac{1}{4 s (\kappa + \sqrt{s})} \nonumber\\
\int \stackrel{(1)}{W_0}({\bf r_{\omega}},{\bf r_{\alpha}};s)
~d\stackrel{(p)}{S}_{\alpha} & & =
-\frac{1}{2 \sqrt{s}^3(\kappa + \sqrt{s})}\times \frac{1}{R_{\omega}}
\nonumber\\
\int \stackrel{(2)}{W_0}({\bf r_{\omega}},{\bf r_{\alpha}};s)
~d\stackrel{(p)}{S}_{\alpha} & & = 
\frac{1}{4 s^2 (\kappa + \sqrt{s})}\times\left[ \frac{1}{R_{\omega}^2} -
\frac{(2\kappa + \sqrt{s})}{2(\kappa + 
\sqrt{s})}\left(\frac{1}{R_{\omega}^2} -
\frac{1}{R_{\omega_1}R_{\omega_2}}\right) \right] \nonumber\\
~ 
\label{equ22bla}
\end{eqnarray}
that are on the other hand the only quantities that we need in
evaluation of the partition functions Eq.\ref{equxi}. This being so
due to the fact that $F({\bf r_{\alpha}},{\bf r_{\beta}};s)$ depends
only on the difference between the coordinates and not on their
absolute values. This property then reverberates through the defining
equations all the way to the partition function Eq.\ref{equxi}.

We now follow the definitions Eqs.\ref{equ8}, \ref{equ9} and
\ref{equ10} and thus as a result obtain a curvature expansion for
$\Xi_0({\cal N})$, $\Xi_1({\cal N})$ and $\Xi_2({\cal N})$. To the
lowest order in the average curvature ${1\over{R_{\omega}}} =
{\textstyle\frac{1}{2}} \left( {1\over{R_{\omega 1}}} + {1\over{R_{\omega 2}}} \right)$, 
and in the limit of long polymers, ${\cal N} \gg 1$, we derive the
following curvature expansion for the case $\kappa > 0$
\begin{eqnarray}
\kern-30pt\Xi_0({\cal N}) &=& V - {{\sqrt{\cal N}}\over{\sqrt{\pi}}}\int 
dS_{\omega} \left[ 1 - 
\sqrt{\pi}\sqrt{\cal N}{1\over{R_{\omega}}} + {{2\cal N}\over{3}} \left( 
{1\over{R^2_{\omega}}} + {1\over{R_{\omega 1}R_{\omega 2}}} \right)  \right] + \dots
\nonumber\\
\Xi_1({\cal N}) &=& {1\over{\kappa^2}}\int dS_{\omega} \left[ 1 - {{4\sqrt{\cal N}}\over
{\sqrt{\pi}}} {1\over{R_{\omega}}} + {{\sqrt{\cal N}}\over{\kappa \sqrt{\pi}}}
\left( {1\over{R^2_{\omega}}} - {1\over{R_{\omega 1}R_{\omega 2}}} 
\right) + {\cal N}{1\over{R_{\omega 1}R_{\omega 2}}} \right] + \dots \nonumber\\
\Xi_2({\cal N}) &=& {4\over{\kappa^3}}\int dS_{\omega} \left[ 1 - {{4\sqrt{\cal N}}\over
{\sqrt{\pi}}} {1\over{R_{\omega}}} + {{2\sqrt{\cal N}}\over{\kappa \sqrt{\pi}}}
\left( {1\over{R^2_{\omega}}} - {1\over{R_{\omega 1}R_{\omega 2}}} 
\right) + {\cal N}{1\over{R_{\omega 1}R_{\omega 2}}} \right] + \dots \nonumber\\
.
\label{equ15}
\end{eqnarray}
where $\omega$ stands for the index of the coordinate over which the
final integration of the $Tr$ operation is carried out, see Eq.\ref{equ10a}.

In the opposite case of $\kappa < 0$ the spectrum of $G_0({\bf r}_{\alpha},{\bf
r}_{\beta};{\cal N})$ has bound states leading to the following
dependence on ${\cal N} \gg 1$ after the inverse Laplace transform
\begin{eqnarray}
\kern-30pt\Xi_0({\cal N}) &=& V + \frac{e^{\kappa^2 {\cal
N}}}{\kappa}\int dS_{\omega} \left[1 - \frac{2}{\kappa}
\frac{1}{R_{\omega}} + {\cal N}\left( \frac{1}{R_{\omega}^2} -
\frac{1}{R_{\omega_1}R_{\omega_2}} \right) \right] + \dots
\nonumber\\
\Xi_1({\cal N}) &=& \frac{4(\kappa^2{\cal N})~e^{\kappa^2{\cal
N}}}{\kappa^2}\int dS_{\omega} \left[ 1 -
\frac{2}{\kappa}\frac{1}{R_{\omega}} +
\frac{1}{\kappa^2}\frac{1}{R_{\omega}^2} + {\cal N}\left( \frac{1}{R_{\omega}^2} -
\frac{1}{R_{\omega_1}R_{\omega_2}} \right) \right] + \dots
\nonumber\\
\Xi_1({\cal N}) &=& \frac{16(\kappa^2{\cal N})^2~e^{\kappa^2{\cal
N}}}{\kappa^3}\int dS_{\omega} \left[ 1 -
\frac{2}{\kappa}\frac{1}{R_{\omega}} +
\frac{1}{\kappa^2}\frac{1}{R_{\omega}^2} + {\cal N}\left( \frac{1}{R_{\omega}^2} -
\frac{1}{R_{\omega_1}R_{\omega_2}} \right) \right] + \dots
\nonumber\\
~
\label{equbound}
\end{eqnarray}
One should not forget at this point that all the above relations were
derived under the restriction that $ \vert \kappa \vert \gg
{{\partial \log{{\cal G}_0({\bf r}_{\alpha},{\bf r}';s)}}\over
{\partial {\bf n}_{\alpha}}}$. This restriction is not particularly
stringent if one does not approach the region $\kappa \cong 0$. 

\section{Results and Discussion}

Taking now equations Eqs.\ref{equ11a} and \ref{equ15} and assuming
that the surface density ${\textstyle\frac{M}{S_{\omega}}}$ and the
volume density ${\textstyle\frac{N}{V}}$ are finite , we end up with
the following form of the free energy in the limit of $\kappa > 0$
\begin{eqnarray}
\beta{\cal F} &\cong& \beta {\cal F}_0 - (N - M)\ln{V} -
M\ln{S_{\omega}} + \nonumber\\
&+& \left({N\over V}\right) \left[ {{\sqrt{\cal N}}\over{\sqrt\pi}} 
\int dS_{\omega}  - {\cal N}\int dS_{\omega} 
{1\over{R_{\omega}}} +
{{2\sqrt{{\cal N}}^3}\over{3\sqrt{\pi}}} \int dS_{\omega} \left( 
{1\over{R^2_{\omega}}} + {1\over{R_{\omega 1}R_{\omega 2}}} \right) \right] + 
\nonumber\\
&+& \left( {M\over{S_{\omega}}}\right) \left[ {{4\sqrt{\cal
N}}\over{\sqrt{\pi}}}\int dS_{\omega}{1\over{R_{\omega}}} -
{{\sqrt{\cal N}}\over{\kappa \sqrt{\pi}}}\int dS_{\omega} \left(
{1\over{R^2_{\omega}}} - {1\over{R_{\omega 1}R_{\omega 2}}}
\right) - {\cal N}\int dS_{\omega} {1\over{R_{\omega 1}R_{\omega 2}}}\right] -
\nonumber\\
&-& {\kappa \over 2}
\left({M\over{S_{\omega}}}\right)^2\left({N\over V}\right)^{-1} \left[ 
\int dS_{\omega} + {{4\sqrt{\cal N}}\over{\sqrt{\pi}}} \int
dS_{\omega}{1\over{R_{\omega}}} - {\cal N}\int dS_{\omega}
{1\over{R_{\omega 1}R_{\omega 2}}}\right] + \dots \nonumber\\
~
\label{equ16}
\end{eqnarray}
$\beta {\cal F}_0$ contains irrelevant constants and combinatorial terms, 
while the second and the third terms describe the removal of $M$ out of total 
of $N$ volume translational degrees of freedom and creation of $M$ surface 
translational degrees of freedom due to the existence of mobile surface 
sliplinks. The nature of the other terms in the free energy is also
straightforwardly discernible (see below).

In the opposite limit of $\kappa < 0$, assuming now  that the two
surface densities ${\textstyle\frac{M}{S_{\omega}}}$ and ${\textstyle\frac{N}{S_{\omega}}}$ are
finite, we derive the following form of the free energy
\begin{eqnarray}
\beta {\cal F} &\cong& \beta{\cal F}_0 - N\ln{S_{\omega}} + \nonumber\\
&+& \left(\frac{N}{S_{\omega}}\right)\left[ - \kappa^2 {\cal N}
 \int dS_{\omega} +  
\frac{2}{\kappa}\int dS_{\omega} \frac{1}{R_{\omega}} - {\cal N}\int
dS_{\omega}\left( \frac{1}{R_{\omega}^2} -
\frac{1}{R_{\omega_1} R_{\omega_2}} \right)\right] -
\nonumber\\
&-& \left(\frac{M}{S_{\omega}}\right) \frac{1}{\kappa^2}\int
dS_{\omega} \frac{1}{R_{\omega}^2} -
{\textstyle{\frac{1}{2}}} \left(\frac{M}{S_{\omega}}\right)^2 \left(
\frac{N}{S_{\omega}} \right)^{-1}\frac{1}{\kappa^2}\int
dS_{\omega} \frac{1}{R_{\omega}^2} + \dots \nonumber\\
~
\label{free2}
\end{eqnarray}
where the second term now simply signifies that in this limit the
chains are mostly adsorbed to the surface as we assumed that there are
no volume interactions between them. 

The above two results should be compared with the canonical form of
the membrane elastic energy
\cite{evans}
\begin{equation}
{\cal F} = {\textstyle\frac{1}{2}} K_c \int dS_{\omega} \left(
\frac{1}{R_{\omega}} - \frac{1}{R_0} \right)^2 + K_G \int dS_{\omega}
\frac{1}{R_{\omega 1}R_{\omega 2}},
\end{equation}
wherefrom one obtains the contribution of the polymer - surface
interactions to the curvature modulus $(K_c)$, spontaneous curvature
radius $(R_0)$, as well as to the modulus of Gaussian curvature $(K_G)$.

On comparison with results of Ref.\cite{ref1}, it is clear that the fourth term
in the free energy Eq.\ref{equ16} corresponds to the statistically
averaged contact excluded volume interaction of the polymer solution
with the deformed boundary, as the limit $\vert \kappa \vert \gg 0$
assumed in deriving this result essentially reduces to the Dirichlet
boundary condition.  We shall referr to this term in the free energy
as the contact term.

The term linear in ${\textstyle\frac{M}{S_{\omega}}}$ is the only one that vanishes
completely for a flat surface and is formally analogous to an
effective polymer adsorption term \cite{ref1}. As it is non - zero
only for a rough surface it stemms from the "in plane bridging" (see
Ref.\cite{ref1}) provided by the segments of the polymers between different
sliplinks. Obviously it tends to curve the surface towards the
polymer rich side and thus acts in opposition to the
contact term.

The last term in Eq.\ref{equ16}, being of the second order in
${\textstyle\frac{M}{S_{\omega}}}$, corresponds to the free energy of chain segments
caught by the sliplinks and thus brought in the immediate vicinity of
the repulsive surface. Its zero and first order curvature terms are
both negative, reflecting the predominant influence of the deminished
entropy over the interaction energy ($2\kappa$ per link) of the chains
meeting at the sliplink confined to the supporting surface. This term
acts to expose a larger area to the polymer rich side, thus tending to
curve the supporting surface away from it in concert with the contact
term. It is also the only term in the energy expansion that does not
effect the curvature modulus at all.

The final rescaling of the curvature modulus and spontaneous curvature
in this limit then assume the form
\begin{equation}
K_c \longrightarrow K_c + {\textstyle\frac{2}{3}}kT\phi\frac{{\cal R}_G^3}{\sqrt{\pi}} - kT
\left( \frac{M}{S_{\omega}}\right) \frac{{\cal R}_G}{\kappa
\sqrt{\pi}} + \dots
\end{equation}
and
\begin{equation}
\frac{K_c}{R_0} \longrightarrow \frac{K_c}{R_0} + kT \phi {\cal R}_G^2\left( 1 +
2\kappa \left({M\over{S_{\omega}}}\right)^2{{\phi^{-2}}\over{{\cal
R}_G}}\right) - kT \left( \frac{M}{S_{\omega}}\right)\frac{4}{\sqrt{\pi}}
{\cal R}_G + \dots,
\end{equation}
where we introduced the radius of gyration, which in appropriate units 
assumes the form ${\cal R}_G = \sqrt{\cal N}$ and the volume
density of polymers $\phi = \frac{N}{V}$.

The free energy of the system in the limit of $\kappa \gg 0$ thus
emerges as being composed of three decoupled terms: the surface energy
of the bulk polymer solution due to the repulsive interaction of the
polymers with the surface, the energy of those polymer chains that
happen to join different sliplinks and lie in the plane of the
unperturbed surface, and the surface energy of sliplinks themselves,
since at the points of crosslinking different segments of the chains
make contact with the (repulsive) surface.  After relaxing the limit
${\textstyle\frac{M}{N}} \ll 1$ we can presume, that only the last two mechanisms
will remain in making a substantial contribution to the total free
energy. Unfortunately the relaxing of this limit makes the whole
calculation much more demanding, probably due to the fact that the
system goes through a surface gelation transition.

In the case of attractive interactions between the chains and the
surface the third term corresponds to the free energy of a polymer
bound to a deformed surface with energy $\kappa$ per surface area. The
first term in the parenthesis is simply the adsorption energy
\cite{eisenriegler}. The term linear in curvature preferrs bending towards the
polymer rich side and thus acts exactly in the opposite direction as
the analogous term in the contact term Eq.\ref{equ16}. This is also
the only term in the whole free energy expression that is linear in
curvature. 

The rescaling of the curvature modulus obviously stemms in
this case from three different mechanisms which interestingly enough
all act in the direction of destabilizing the curvature modulus. The
first contribution, being linear in $\frac{N}{S_{\omega}}$ comes from
the adsorption energy and is of the same form as in the case when no
pinning to sliplinks is present \cite{ref1}. The second and third
contributions, being linear and quadratic in $\frac{M}{S_{\omega}}$
respectively, are obviously due to the presence of surface - bound
sliplinks. The linear term is due to the energy difference in the
pinning at the sliplink as compared to the 'soft' adsorption while the
third term is due to the energy difference of an adsorbed portion of
the chain when it is pinned by two sliplinks, at the beginning and at
the end, and when it is 'softly' adsorbed.

The final rescaling of the elastic properties of the ``dressed''
membrane can be in this limit cast into the form
\begin{equation}
K_c \longrightarrow K_c - 2\left(\frac{N}{S_{\omega}}\right){\cal
R}_G^2 - 2\left(\frac{M}{S_{\omega}}\right)\kappa^{-2}\left( 1 +
{\textstyle\frac{1}{2}}\left(\frac{M}{S_{\omega}}\right)^2\left(\frac{N}{S_{\omega}}
\right)^{-2}\right) + \dots
\label{prut}
\end{equation}
and
\begin{equation}
\frac{K_c}{R_0} \longrightarrow \frac{K_c}{R_0} +
\left(\frac{N}{S_{\omega}}\right)\frac{2}{\kappa} +\dots.
\end{equation}

The modifications in the Gaussian modulus are important because they
extend only over the area occupied by polymers and are thus position
dependent (otherwise they would make no contribution to the equations
determining the shape of the ``dressed'' membrane). In the case of
repulsive polymer - membrane interactions, {\sl i.e.} $\kappa > 0$ we
obtain the following form for the change in the Gaussian curvature
modulus
\begin{equation}
K_G \longrightarrow K_G + \phi{{2{\cal R}_G^3}\over{3\sqrt{\pi}}} -
\left( {M\over{S_{\omega}}}\right)
\left[{\cal R}_G^2 - {{{\cal R}_G}\over{\kappa \sqrt{\pi}}}\right] + 
{\kappa \over 2}\left({M\over{S_{\omega}}}\right)^2\phi^{-1}{\cal R}_G^2 + \dots .
\label{pos}
\end{equation}
Since the signs of different contributions in this case are not
uniform there is no general statement that one could make regarding
the proliferation of handles.  

In the opposite case of attractive polymer - membrane interactions,
$\kappa < 0$, the Gaussian curvature modulus has a very simple
dependence on the polymer parameters
\begin{equation}
K_G \longrightarrow K_G + \left(\frac{N}{S_{\omega}}\right){\cal R}_G^2 +
\dots .
\label{neg}
\end{equation}
that in conjunction with Eq.\ref{prut} favours the proliferation of
handles. The last result is interesting also from the point of view of
recent analysis of the modifications in the Gaussian curvature modulus
of partially polymerized surfactant membranes \cite{kozlov}. Clearly
for large negative values of $\kappa$ our problem is isomorphous to
the problem of polymers embedded in the membrane. In case there are no
interactions between embedded polymer chains the result derived by
Kozlov and Helfrich \cite{kozlov} for this particular model system
reduces exactly to Eq.\ref{neg}.

\vfill
\newpage
\noindent
{\bf Figure Captions}

\vskip 1 cm

\noindent
{\bf Fig.1}~Schematic representation of the model system. Polymers are
pinned to the supporting surface ${\bf r}_{\alpha}$ with sliplinks of
functionality four that are mobile along the surface. The interaction
of the chains between the sliplinks with the surface is short ranged
and described with the phenomenological parameter $\kappa$.

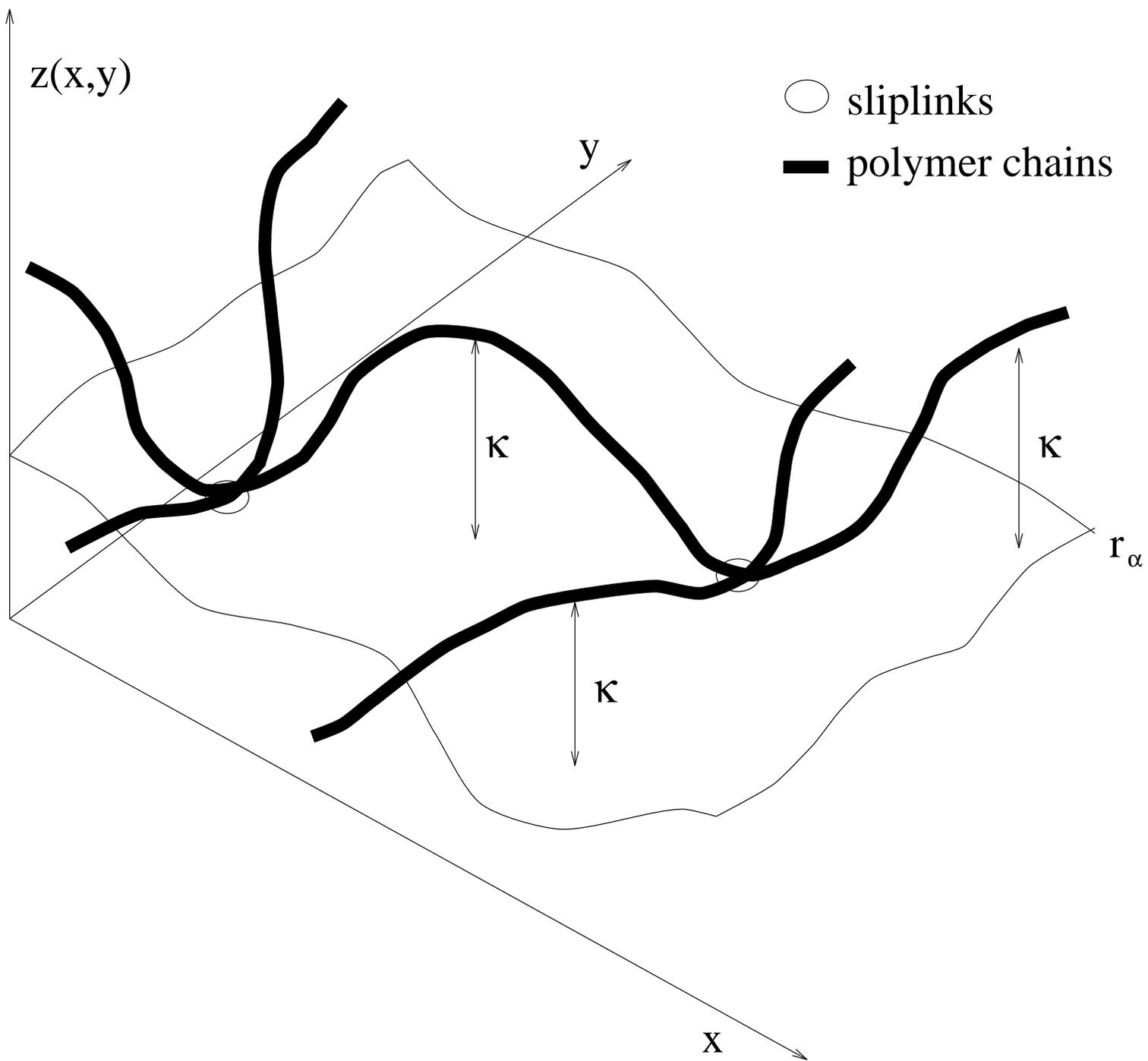
\end{document}